\documentclass[aps,prb,twocolumn,floats,showpacs]{revtex4}
\usepackage{graphicx}%
\usepackage[]{epsfig}
\usepackage{dcolumn}
\usepackage{amsmath}
\usepackage{amssymb}
\usepackage{bbold}
\usepackage{latexsym}

\newcommand{\la}{\langle}
\newcommand{\ra}{\rangle}

\renewcommand{\vr}{\mathbf{r}}

\begin{document}
\bibliographystyle{apsrev}
\title{On the scaling approach to electron-electron
interactions in a chaotic quantum dot}
\author{Shaffique Adam, Piet W.\ Brouwer, and Prashant Sharma}
\affiliation{Laboratory of Atomic and Solid State Physics,
Cornell University, Ithaca, NY 14853}
\date{\today}
\begin{abstract}
  A scaling theory is used to study the low energy physics of 
  electron-electron interactions in a double quantum dot. We show
  that the fact that electrons are delocalized over two quantum
  dots does not affect the instability criterion for the
  description of electron-electron interactions in terms of a
  ``universal interaction Hamiltonian''.
\end{abstract}


\maketitle

The statistical distribution of single-particle energy levels 
and wavefunctions in a 
chaotic quantum dot or disordered metal particle is described by
random matrix theory.\cite{kn:alhassid2000b,kn:guhr1998}
The validity of random matrix theory as a statistical description
of energy levels and wavefunctions follows from
the existence of a large parameter, the dimensionless conductance
$g$ of the metal grain or the quantum dot.\cite{kn:altshuler1986}
(The dimensionless conductance is the ratio of the Thouless energy
$E_{\rm T}$ and the mean level spacing $\Delta$.)
The same large parameter $g$ allows for a consistent and simple
description of electron-electron interactions in quantum dots and
metal grains, by means of the ``universal interaction Hamiltonian'',
which was proposed by Kurland, Aleiner, and 
Altshuler\cite{kn:kurland2000} (see also Ref.\ \onlinecite{kn:aleiner2002}).
According to Ref.\ \onlinecite{kn:kurland2000}, 
to leading order in $g$, the only relevant 
contributions to the interaction Hamiltonian are the capacitive 
charging energy, the long-range exchange interaction, and the
``Cooper-channel'' interaction, which is responsible for the
superconducting instability.

The justification for the ``universal interaction
Hamiltonian'' follows from the statistics of wavefunctions 
$\phi_{\alpha}$ in disordered
metal grains or chaotic quantum dots. Wavefunctions determine the 
matrix elements of the electron-electron interaction,
\begin{eqnarray}
  V_{\alpha\beta\gamma\delta} &=& 
  \int d\vr_1 d\vr_2 \phi_{\alpha}(\vr_1)^*
  \phi_{\beta}(\vr_2)^*
  \nonumber \\ && \mbox{} \times
  V(\vr_1,\vr_2) 
  \phi_{\gamma}(\vr_2)
  \phi_{\delta}(\vr_1).
\end{eqnarray}
The absence of (long-range) wavefunction correlations in
chaotic quantum dots causes interaction matrix elements to be
self-averaging. Most averages are zero, except averages of
``diagonal''
interaction matrix elements $V_{\alpha\beta\gamma\delta}$ where the 
wavefunction indices
coincide pairwise. Replacing interaction matrix elements
$V_{\alpha\beta\gamma\delta}$ by their ensemble average 
$\langle V_{\alpha\beta\gamma\delta}\rangle$, only the charging energy,
exchange coupling, and Cooper channel interaction remain, thus
leading to the ``universal interaction Hamiltonian''.
Small non-universal corrections to the interaction Hamiltonian
follow from residual wavefunction 
correlations in disordered metal grains or quantum
dots, which cause small fluctuations of the interaction matrix elements
$V_{\alpha\beta\gamma\delta}$ around their average. Typically, these
fluctuations are a factor $1/g$ smaller than the
diagonal matrix elements.

Although the off-diagonal interaction matrix elements are a factor
$1/g$ smaller than the diagonal elements, they are many, and it is
legitimate to ask what their role is. This question was addressed by
Murthy and coworkers using a renormalization-group approach, in a
series of papers.\cite{kn:murthy2002,kn:murthy2003,kn:murthy2003b}
These authors assumed Fermi Liquid interactions on time scales
shorter than $\hbar/E_{\rm T}$, and used random matrix theory to
describe electron dynamics on time scales beyond $\hbar/E_{\rm
T}$. Successively integrating out states with highest energy, they
found that the ``universal interaction Hamiltonian'' is stable for 
repulsive
Fermi-liquid interactions and for weak attractive Fermi-liquid
interactions, whereas an instability occurs when the attraction is
sufficiently strong. Remarkably, Murthy {\em et al.}  found that the
critical attraction strength is a factor $2 \ln 2$ smaller than the
attraction strength corresponding to the Pomeranchuk instability in
the bulk Fermi liquid, thus creating a parameter regime where the bulk
system is stable, whereas the finite-sized system is
not.\cite{kn:murthy2002,kn:murthy2003,kn:murthy2003b}

A renormalization-group treatment of interactions in chaotic quantum
dots requires knowledge of how (non-universal) wavefunction
correlations depend on the energy difference between the wavefunctions
involved. The answer to this question depends on the detailed shape of
the quantum dot and is different for diffusive\cite{kn:mirlin2000} and
ballistic\cite{kn:aleiner1996} electron dynamics (see also Ref.\ 
\onlinecite{kn:aleiner2002}).  Murthy {\em et al.}\ bypass this
problem by using the eigenfunction correlations of a $g$-dimensional
random matrix for all wavefunctions with energy within $E_{\rm T}/2$
from the Fermi level, treating wavefunctions at larger energies as
plane waves. Whereas the use of random matrix theory is justified for
energies far below $E_{\rm T}$ only, it cannot be used to describe
non-universal wavefunction statistics near the Thouless energy.
Similarly, residual wavefunction correlations will persist for
energies above $E_{\rm T}$, which are not accounted for in Refs.\ 
\onlinecite{kn:murthy2002,kn:murthy2003,kn:murthy2003b}. A correct
treatment of wavefunction correlations around $E_{\rm T}$ is important
for the renormalization group approach, since most of the
renormalization of the interaction parameters takes place around that
energy.

\begin{figure}[t]
\epsfxsize=0.88\hsize
\epsffile{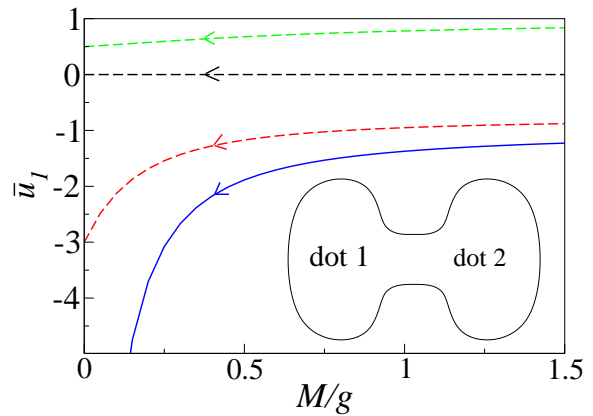}
\caption{\label{fig:1}
Renormalization group flow of the effective dipolar coupling $\bar
u_1$ as a 
function of the cutoff $M$.  The solid curve corresponds to the
critical dipolar interaction strength $u_1 = -1$. The other 
curves are (from bottom to top) for $u_1 = -0.75$, $0$ and $1$. 
Inset: schematic drawing of the double quantum dot.}
\end{figure}

In this communication we apply the renormalization group scheme to the
special case of a ``double quantum dot'', see Fig.\ \ref{fig:1},
inset. The
``double quantum dot'' consists of two quantum dots of roughly equal size
coupled via a point contact with dimensionless conductance $g/2 \gg
1$. The Thouless energy $E_{\rm T}$ of the double dot system is equal
to $g \Delta$, where $\Delta$ is the double-dot level spacing (which
is half the single-dot level spacing). 
The dimensionless conductances of the two individual quantum
dots are assumed to be much larger than $g$, so that random 
matrix theory and the
``universal interaction Hamiltonian'' can be used to describe
wavefunctions and interactions in each of the dots separately. The
advantage of the double dot geometry is that wavefunction correlations
for energy differences near $E_{\rm T}$ can be calculated in
detail, so that no approximations need to
be made upon constructing the renormalization group for the
electron-electron interactions. The analogy between the double dot
system studied here and the single quantum dot studied by Murthy 
{\em et al.}\ is that in the double dot electrons
are confined to one dot for times well below
$\hbar/E_{\rm T}$, but not for larger times, whereas in the
ballistic quantum dot studied in 
Refs.\ \onlinecite{kn:murthy2002,kn:murthy2003,kn:murthy2003b} 
they have a well-defined momentum for 
times below $\hbar/E_{\rm T}$, but not for longer times. 

Our main finding, to be 
elaborated below, is that, once the correct non-universal
wavefunction correlations
near $E_{\rm T}$ are taken into account, the instability of the
universal Hamiltonian occurs at precisely the same 
interaction strength as the instability of the double-dot system
without point contact between the dots. Although
this conclusion is reached for one specific geometry
only, 
the structure of our calculation leads us to 
expect that the same is true for the more general
Pomeranchuk-type instabilities studied by Murthy {\em et al.}
In other words, we expect that the fact that 
Refs.\ \onlinecite{kn:murthy2002,kn:murthy2003,kn:murthy2003b} 
find a parameter regime where the bulk Fermi Liquid is stable 
whereas the
finite-sized system is not is an artifact of the use of
random matrix theory to describe wavefunction statistics up to
a distance $E_{\rm T}/2$ from the Fermi level.

We now describe the details of our calculation. For technical
convenience,
we consider a double quantum dot with spinless electrons and
with broken time-reversal symmetry. Using random matrix theory
to describe each of the dots separately, the non-interacting 
part of the Hamiltonian for the double quantum dot 
reads\cite{foot1}
\begin{equation}
  H = \left( \begin{array}{cc} H_1 & 0 \\ 0 & H_2 \end{array}
  \right)
  + \sqrt{\frac{g}{8 N}} H_{12},
  \label{eq:H}
\end{equation}
where $H_1$ and $H_2$ are $N \times N$ hermitian matrices modeling the
Hamiltonians of the quantum dots without point contact and $H_{12}$
is a $2N \times 2N$ hermitian matrix modeling the point contact
connecting the two quantum dots. The elements of $H_1$, $H_2$, and
$H_{12}$ are complex numbers taken from independent and identical
Gaussian distributions. The rows and columns of $H$ are labeled by 
roman numbers $k=1,\ldots,2N$, where $k=1,\ldots,N$ and 
$k=N+1,\ldots,2N$ correspond to the left and right dots, respectively.
Eigenvalues of $H$ are denoted $\varepsilon_{\alpha}$, the
corresponding eigenvector being written $\phi_{\alpha}(k)$.
The size $N$ of the random matrices $H_1$ and $H_2$ 
is of the order of the dimensionless
conductance of the individual quantum dots and is taken to infinity 
at the end of the calculation.

The interaction Hamiltonian has the form\cite{foot2}
\begin{equation}
  H_{\rm int} = \frac{1}{2} U_0 (\hat n_1 + \hat n_2)^2 + 
  \frac{1}{2} U_1 (\hat n_1 -  \hat n_2)^2, \label{eq:Hint}
\end{equation}
where $\hat n_1$ and $\hat n_2$ are operators for the number of
electrons in the two individual quantum dots. The first term in Eq.\ 
(\ref{eq:Hint}) corresponds to a ``charging energy'' for the double
dot system, whereas the second term in Eq.\ (\ref{eq:Hint}) represents
a dipolar interaction.  Upon changing to the basis of eigenstates of
the non-interacting Hamiltonian $H$, the total Hamiltonian reads
\begin{equation}
  {\cal H} = \sum_{\alpha} \varepsilon_{\alpha} \hat \psi^{\dagger}_{\alpha}
  \hat \psi^{\vphantom{\dagger}}_{\alpha} +
  \frac{1}{2}
  \sum_{\alpha\beta\gamma\delta} V_{\alpha\beta\gamma\delta}
  \hat \psi^{\dagger}_{\alpha} \hat \psi^{\dagger}_{\beta}
  \hat \psi^{\vphantom{\dagger}}_{\gamma} \hat 
  \psi^{\vphantom{\dagger}}_{\delta},
\end{equation}
where $\hat \psi^{\dagger}_{\alpha}$ and $\hat
\psi^{\vphantom{\dagger}}_{\alpha}$ are creation and annihilation 
operators for an electron in eigenstate $\alpha$ of the non-interacting
Hamiltonian $H$ and
\begin{eqnarray}
  \label{eq:vabcd}
  V_{\alpha\beta\gamma\delta}&=&
  \sum_{k,l} (U_0 +
   U_1 \sigma_k \sigma_l)
  \phi^*_\alpha(k)\phi^*_\beta(l)
  \phi_\gamma(l)\phi_\delta(k),
\end{eqnarray}
where $\sigma_k = 1$ for $k=1,\ldots,N$ and $\sigma_k = -1$ for
$k=N+1,\ldots,2N$.

For large $g$, wavefunction elements $\phi_{\alpha}(k)$
are independently distributed Gaussian random numbers,
\begin{equation}
  \langle \phi_{\alpha}(k) \phi_{\beta}(l) \rangle
  = \frac{1}{2N} \delta_{kl} \delta_{\mu \nu},
  \label{eq:Gauss}
\end{equation}
corrections to Eq.\ (\ref{eq:Gauss}) being of order $1/g$. As a
result, interaction matrix elements are self averaging;
fluctuations are of relative order $1/g$. Only
diagonal elements have a 
nonzero average,
\begin{equation}
  \langle V_{\alpha \beta \gamma \delta} \rangle
  = U_0 \delta_{\alpha\delta} \delta_{\beta\gamma}.
\end{equation}
Replacing the interaction matrix elements by their average, we find that
interactions are described by the reduced interaction Hamiltonian
\begin{equation}
  H_{\rm int} = \frac{U_0}{2}\hat n^2, \label{eq:HintU}
\end{equation}
where $\hat n = \hat n_1 + \hat n_2$ is the total number of
electrons in the double quantum dot. Notice that, in comparison
to Eq.\ (\ref{eq:Hint}), the dipolar interaction
has disappeared because the electron wavefunctions are delocalized 
over of the entire double-dot system.
Equation (\ref{eq:HintU}) is 
the equivalent of the ``universal interaction
Hamiltonian'' for the double quantum dot; the disappearance of the
dipolar interaction is
the double-dot counterpart of the disappearance of all non-zero-mode
Fermi Liquid interactions in the original construction of the
``universal interaction
Hamiltonian''.\cite{kn:kurland2000,kn:aleiner2002}

\begin{figure}[t]
\epsfxsize=0.8\hsize
\vspace{0.5cm}
\epsffile{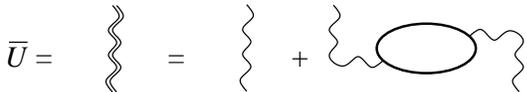}
\caption{\label{fig:diagr}
Diagrammatic representation of the effective interaction
$\bar U$, to second order in perturbation theory.}
\end{figure}

In order to study the importance of the many residual interaction matrix
elements that are of order $1/g$, we perform a renormalization group
analysis, following Refs.\
\onlinecite{kn:murthy2002,kn:murthy2003,kn:murthy2003b} (see also
Ref.\ \onlinecite{kn:shankar1994}). 
This analysis is reminiscent
of Anderson's ``poor man's'' treatment of the Kondo
problem:\cite{kn:anderson1970} In order to find the effective
interaction for electrons at the Fermi level $\varepsilon_F$, 
one successively integrates out
states at energies far away from $\varepsilon_F$.
Writing the cut-off energy
as $M \Delta/2$, we calculate the change of the
effective interaction parameters $\bar U_0$ and $\bar U_1$ 
upon changing $M$ to
$M'<M$ within second order perturbation theory, see Fig.\
\ref{fig:diagr}
\begin{eqnarray}
  \bar U_0(M') - \bar U_0(M)
   &=& \bar U_{0}^2 
  \sum_{k,l} \sum_{\mu \nu}\!'
  \frac{n_F(\varepsilon_{\mu}) -
  n_F(\varepsilon_{\nu})}{\varepsilon_{\mu} - \varepsilon_{\nu}}
  \nonumber \\ && \mbox{} \times
  \phi_{\mu}^*(k) \phi_{\mu}(l) \phi_{\nu}^*(l) \phi_{\nu}(k), 
  \label{eq:upert0}
  \\
  \bar U_1(M') - \bar U_1(M)
   &=& \bar U_{1}^2 
  \sum_{k,l}   \sigma_k \sigma_l  \sum_{\mu \nu}\!'
  \frac{
  n_F(\varepsilon_{\mu}) -
  n_F(\varepsilon_{\nu})}{\varepsilon_{\mu} - \varepsilon_{\nu}}
  \nonumber \\ && \mbox{} \times
  \phi_{\mu}^*(k) \phi_{\mu}(l) \phi_{\nu}^*(l) \phi_{\nu}(k).
  \label{eq:upert1}
\end{eqnarray}
where $n_F(\varepsilon)$ is the Fermi function and the sum over
intermediate states is such that only states $\mu$ and $\nu$ with at
least one of the energies $\varepsilon_{\mu}$ or $\varepsilon_{\nu}$
in the cut-off region $M' \Delta/2 < |\varepsilon - \varepsilon_F| < M
\Delta/2$ are to be included. We omitted exchange contributions to the
effective interaction, which are unimportant for the large-$g$ limit
we consider here.  We follow Refs.\ 
\onlinecite{kn:murthy2002,kn:murthy2003,kn:murthy2003b} in replacing
the product of the eigenfunctions $\phi_{\mu}$ and $\phi_{\nu}$ of the
intermediate energies $\varepsilon_{\mu}$ and $\varepsilon_{\nu}$ in
Eqs.\ (\ref{eq:upert0}) and (\ref{eq:upert1}) by its ensemble average.
However, we deviate from Refs.\ 
\onlinecite{kn:murthy2002,kn:murthy2003,kn:murthy2003b} in keeping the
precise dependence of the ensemble average of the wavefunctions
$\phi_{\mu}$ and $\phi_{\nu}$ on the energy difference
$\varepsilon_{\mu} - \varepsilon_{\nu}$. Repeating the analysis of
Ref.\ \onlinecite{kn:adam2002} for the Hamiltonian (\ref{eq:H}), the
relevant wavefunction average is found to be (see also Ref.\ 
\onlinecite{kn:tschersich2000})
\begin{eqnarray}
  \la\phi^*_\nu(k)\phi_\mu(k)\phi^*_\mu(l)\phi_\nu(l)
  \ra
  &=&
  \frac{g\sigma_{k} \sigma_{l} \Delta^2}{4 N^2[g^2 \Delta^2
  + \pi^2(\varepsilon_{\mu}
  - \varepsilon_{\nu})^2]}
  \nonumber \\ && \mbox{}
  + \frac{\delta_{\mu \nu} + \delta_{kl}}{4 N^2}
  - \frac{1+\sigma_k\sigma_l}{8 N^3}.
  \nonumber \\
  \label{eq:cumulant}
\end{eqnarray}
Substituting this into Eqs.~(\ref{eq:upert0}) and (\ref{eq:upert1}), 
we find that only the non-universal
first term on the r.h.s.\ of Eq.\ (\ref{eq:cumulant}) contributes
in the limit $N \to \infty$.
Changing to dimensionless interaction parameters
$\bar u_j = \bar U_j/\Delta$, $j=0,1$, and replacing the difference 
equations (\ref{eq:upert0}) and (\ref{eq:upert1}) by a differential
equation, one thus finds
%
\begin{eqnarray} 
 \frac{d \bar u_0}{dM}&=&0,\label{eq:rg-7a} \\
\frac{d \bar u_1}{dM}&=& \frac{\bar u_1^2}{g}\;
\ln\left(\frac{4 g^2 + M^2 \pi^2}{g^2 + M^2 \pi^2}\right).
  \label{eq:u1flow}
\end{eqnarray}
%
The flow equations (\ref{eq:rg-7a}) and (\ref{eq:u1flow})
are solved with the boundary
conditions $\bar u_j \to u_j = U_j/\Delta$, $j=0,1$, if $M \to
\infty$. Integrating Eqs.\ (\ref{eq:rg-7a}) and (\ref{eq:u1flow}), 
one finds that
$u_0$ does not flow, $\bar u_0(M) = u_0$ for all $M$, whereas
\begin{eqnarray}\label{eq:scaling}
  \bar u_1(M) &=&
  \left[\frac{1}{u_1} + 1 
  - \frac{4}{\pi} \arctan \frac{M \pi}{2 g} 
  + \frac{2}{\pi} \arctan \frac{M \pi}{g}
  \right. \nonumber \\ && \left. \mbox{}
  - \frac{M}{g} 
  \ln \left(\frac{4 g^2 + M^2 \pi^2}{g^2 + M^2 \pi^2} \right)
  \right]^{-1}.
\end{eqnarray}
In Fig.\ \ref{fig:1}, the solution of the flow equation for the
effective dipolar interaction $\bar u_1$ 
is shown for various values of the unrenormalized interaction
$u_1$. 

In order to address the stability of the ``universal interaction
Hamiltonian'', we analyze Eq.\ (\ref{eq:scaling}) in the limit 
$M \downarrow 0$,
\begin{equation}
  \bar u_1(M)= \left( \frac{1}{u_1} + 1 - \frac{2 M}{g} \ln 2 \right)^{-1},\ \
  \mbox{for $M \downarrow 0$}.
  \label{eq:scaling1}
\end{equation}
For $u_1 > -1$, the effective interaction strength $\bar u_1$ 
remains bounded as $M \downarrow 0$. This implies that the
corresponding interaction matrix elements at the Fermi level remain
of order $1/g$, justifying the use of the ``universal interaction
Hamiltonian'' for those values of the dipolar interaction. 
It is only for the
critical dipolar attraction strength $u_1 = -1$ that $\bar u_1$ diverges
upon taking the cut-off energy $M \Delta$ to zero. This is 
precisely at the same interaction strength as the location of the
instability in the absence of inter-dot tunneling.

\begin{figure}[t]
\epsfxsize=0.8\hsize
\epsffile{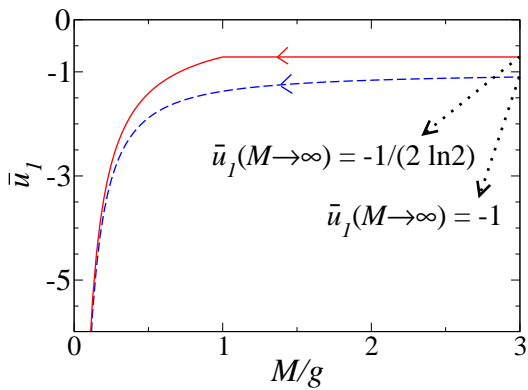}
\caption{\label{fig:2}
Comparison of exact renormalization group flow (dashed)
and the flow according the calculation scheme of Refs.\ 
\onlinecite{kn:murthy2002,kn:murthy2003,kn:murthy2003b} (solid). 
Flows
are shown for the critical value of the dipolar interaction
strength $u_1$.}
\end{figure}

The renormalization approach of Murthy {\em et al.}\ differs from
ours in two respects. First, in Refs.\
\onlinecite{kn:murthy2002,kn:murthy2003,kn:murthy2003b}
there is no flow of the interaction
parameters for $M > g$. Second, in order to describe the flow
for $M < g$, Murthy {\em et al.} replace the eigenfunction
average (\ref{eq:cumulant}) by the average of eigenfunctions of 
a random matrix of size $g$,
\begin{eqnarray}
  \la\phi^*_\nu(k)\phi_\mu(k)\phi^*_\mu(l)\phi_\nu(l)
  \ra &=& \frac{\delta_{\mu \nu} + \delta_{kl}}{g^2}
  - \frac{1}{g^3}.
  \label{eq:cumulant2}
\end{eqnarray} 
One then obtains the following flow equations for the effective
dipolar interaction strength 
$\bar u_1$:
\begin{subequations} \label{eq:rg-7b}
\begin{eqnarray}
  \frac{d \bar u_1}{d M} &=& \lefteqn{0}
  \hphantom{\frac{4 u_1^2 \ln 2}{g}}\ \
  \mbox{if $M > g$}, \\
  \frac{d \bar u_1}{d M} &=& \frac{2 \bar u_1^2 \ln 2}{g}\ \
  \mbox{if $M < g$}.
\end{eqnarray}
\end{subequations}%
The flow equations (\ref{eq:rg-7b}) agree with the exact flow
equations for the double dot system only for $M \ll g$ and $M \gg g$,
but not for the intermediate range $M \sim g$. The solution of
the erroneous flow equations (\ref{eq:rg-7b}) is
\begin{equation}
  \bar u_1(M) = \left\{ \begin{array}{ll} u_1 &
  \mbox{if $M > g$}, \\
  \left[u_1^{-1} + 2 (1 - M/g)  \ln 2 \right]^{-1}
  & \mbox{if $M < g$}.
  \end{array} \right.
  \label{eq:scaling2}
\end{equation}
One verifies that in this calculation scheme, the ``universal
interaction Hamiltonian'' is stable for $u_1 > -1/2\ln2$ only, so that
there is a range of dipolar interaction strengths $-1 < u_1 <-1/2 \ln
2$ for which the separate dots are stable against the formation of a
dipolar charge distribution, whereas the coupled dots are not. Our
exact calculation shows that such a result is incorrect.  In Fig.\ 
\ref{fig:2} we compare the flow of Eq.\ (\ref{eq:scaling2}) and the
exact flow of Eq.\ (\ref{eq:scaling}) for the critical value of $u_1$.
Although it is only in the range $M \sim g$ that the renormalization
group flow of Refs.\ 
\onlinecite{kn:murthy2002,kn:murthy2003,kn:murthy2003b} and the exact
flow for the double dot differ, the flow in the range $M \sim g$ is
crucial in determining the value of the interactions at which the
``universal interaction Hamiltonian'' becomes unstable.

Before concluding, we would like to make three remarks about the
renormalization group calculation presented here. First, the one-loop
renormalization-group result (\ref{eq:rg-7b}) is exact in the
large-$g$ limit. This follows from the same arguments as used to
establish the validity of one-loop renormalization group in the work
of Murthy {\em et al.}\cite{kn:murthy2003}. Second, in the exact
calculation performed here all flow of interaction parameters arises
from the first, non-universal term in the wavefunction correlator
(\ref{eq:cumulant}), which is off-diagonal in the wavefunction
indices. This is opposite to the calculation of Refs.\ 
\onlinecite{kn:murthy2002,kn:murthy2003,kn:murthy2003b}, where the
flow arises from a universal and diagonal wavefunction correlator.
Third, mathematically, the fact that the instability of the
``universal interaction Hamiltonian'' occurs precisely at $u_1=-1$ is
a consequence of the Lorentzian energy dependence of the first,
non-universal term in the wavefunction correlator (\ref{eq:cumulant}).
A Lorentzian is generic for non-universal wavefunction correlations in
both diffusive and ballistic quantum
dots,\cite{kn:mirlin2000,kn:aleiner1996} for which $\sigma_k$ and $g$
in Eq.\ (\ref{eq:cumulant}) are replaced by eigenfunctions and
eigenvalues of the diffusion operator or the Perron-Frobenius
operator, respectively, see, {\em e.g.,} Ref.\ 
\onlinecite{kn:aleiner2002}.  It is because of this similarity that we
believe that our calculational scheme, including our result for the
critical interaction strength, extends to the general case.

We thank G.\ Murthy, R.\ Shankar, and H.\ Mathur for correspondence.
This work was supported by the Cornell Center for Nanoscale Systems
under NSF grant no.\ EEC-0117770, by the NSF under grant no.\ DMR
0086509, and by the Packard foundation.
%


\begin{thebibliography}{14}
\bibitem{foot2}
{We have omitted
an interaction term proportional to $(\hat n_1 + \hat n_2)(\hat n_1 -
\hat n_2)$, which explicitly breaks the symmetry between the two
dots. This interaction is found not to flow upon application 
of the renormalization group scheme.}
\end{thebibliography}
\end{document}